\documentclass[%
 reprint,
 amsmath,amssymb,
 aps,
]{revtex4-1}

\usepackage{graphicx}
\usepackage{dcolumn}
\usepackage{bm}

\newcommand{\bea}{\begin{eqnarray}\displaystyle}
\newcommand{\eea}{\end{eqnarray}}
\newcommand{\nn}{\nonumber}

\newcommand{\figref}[1]{Fig.~\protect\ref{#1}}

{\setlength{\fboxsep}{15pt}
\setlength{\mylength}{\linewidth}%
\addtolength{\mylength}{-2\fboxsep}%
\addtolength{\mylength}{-2\fboxrule}%
\Sbox
\minipage{\mylength}%
\setlength{\abovedisplayskip}{0pt}%
\setlength{\belowdisplayskip}{0pt}%
\equation}%
{\endequation\endminipage\endSbox
\[\fbox{\TheSbox}\]}
\begin{document}


\title{M-strings and Transverse Orbifold}
\author{Khurram Shabbir}

 \vskip 2cm
\affiliation{Department of Mathematics, Government College University, Lahore, Pakistan}


\date{\today}

\begin{abstract}
We discuss the partition function of a single M5-brane on a circle with transverse orbifold of ADE type and show that the modes captured by the partition function are those of the tensor multiplet and the thee form field. We show that the bound states of M-strings corresponding to pair of simple roots appear, for all ADE, only when the momentum on the circle is turned on. 
\end{abstract}

\maketitle

\tableofcontents

\section{Introduction}
In this paper we study the partition function of an M5-brane on a circle with transverse space $S^1\times \mathbb{C}^2/\Gamma$ in the limit the circle size becomes large. This brane configuration has a dual Calabi-Yau geometry which is toric for the case when $\Gamma=\mathbb{Z}_{k}$ i.e., $A$ type orbifold \cite{OMS,Hohenegger}. For the A-type orbifold the partition function can be determined using the refined topological vertex. Interestingly in the toric case $\Gamma=\mathbb{Z}_{k}$ this configuration is dual to the configuration in which we can get rid of the transverse orbifold at the expense of having $k$ M5-branes wrapped on the circle. This duality can be achieved by going to string theory and using T-duality which maps $A$ type orbifold to NS5-branes. As was discussed in \cite{CLST,Hohenegger:2015btj} this can also be understood in terms of T-duality of little string theory.
\begin{figure}[h]
  \centering
  \includegraphics[width=3.5in]{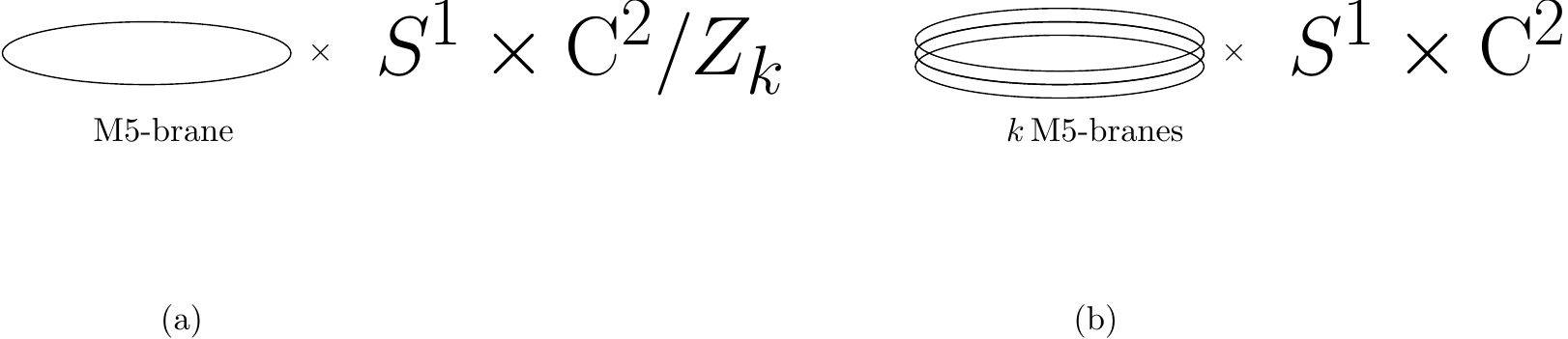}
  \caption{Dual brane configurations.}\label{web1}
\end{figure}

In the case when the transverse space was just $\mathbb{R}^5$ it was shown in \cite{Lockhart:2012vp} that refined topological string partition function of a certain toric elliptic Calabi-Yau threefold captures the field content completely which was just the circle reduction of the $(2,0)$ tensor multiplet. We discuss in detail the case when the space transverse to the M5-brane is $S^{1}\times \mathbb{C}^2/\Gamma_{ADE}$, where $\Gamma_{ADE}$ is the ADE discrete subgroup of $SU(2)$. We show that in the case $\Gamma_{ADE}=\mathbb{Z}_{k}$ there is corresponding toric elliptically fibered CY3fold whose refined topological string partition function gives the modes of the $(2,0)$ tensor multiplet as well as the modes coming of the M-theory three form field reduced on the circle and the orbifold. We also show that for all ADE orbifolds M-strings do not form bound states unless the momentum is turned on the circle on which the M5-brane is wrapped.

This paper is organized as follows. In section 2 we discuss the Calabi-Yau geometries dual to brane configuration shown in \figref{web1} and give the corresponding partition function for the $A$ type orbifold. In this section we also give the dual form of the partition function coming from its realization as $\chi_y$ genus of product of Hilbert scheme of points on $\mathbb{C}^2$. In section 3 we discuss the mode expansion of the refined partition functions and show that they capture the tensor multiplet modes as well as those coming from the three form field. In section 4 we generalize the A type partition function to $D$ and $E$ type and discuss the appearance of bound states when momentum is turned on. In section 4 we discuss our conclusions and some work in progress.

After this paper appeared in the arXiv we learned that discussion in Section II-B overlaps the discussion in \cite{Benvenuti:2016dcs}. 

\section{M5-Brane Configuration and its Partition Function}

\begin{figure}[h]
  \centering
  \includegraphics[width=2in]{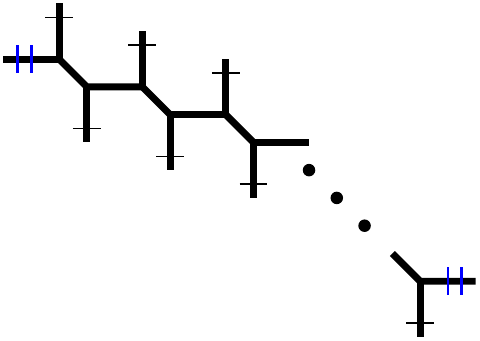}
  \caption{(a) $k$ M5-branes on a circle with transverse space $S^{1}\times \mathbb{C}^{2}/\Gamma$. (b) The dual Calabi-Yau geometry after resolving the singularities of the transverse orbifold.}\label{web2}
\end{figure}

In this section we will discuss the calculation of the partition function of the a single M5-brane with transverse space $S^{1}\times \mathbb{C}^{2}/\mathbb{Z}_{k}$. Going down to type IIB and using T-duality this brane/geometry configuration can be mapped to an intersecting system of a single D5-brane and $k$ NS5-branes. The $SL(2,\mathbb{Z})$ symmetry of type IIB can then map $1\mbox{D5}+k\,\mbox{NS-5}$ to $1\,\mbox{NS-5}+k\,\mbox{D5}$ which can then be lifted back to M-theory to get $k$ M5-branes with transverse space $S^1\times \mathbb{C}^2$ establishing the duality of \figref{web1}. If we resolve the D5-brane/NS5-brane intersections we obtain a brane web shown in \figref{web2}. 

As is well known the $(p,q)$ 5-brane webs are dual to Calabi-Yau threefolds \cite{LV} and in this case there is a double elliptically fibered Calabi-Yau threefolds dual to the above brane configuration \cite{Hohenegger:2015btj, CLST}. This threefold is actually a $\mathbb{Z}_{k}$ orbifold of the threefold corresponding to $k=1$. The $k=1$ threefold is a double elliptic fibration over the complex plane with a single $I_{0}$ fiber of both fibrations. In the limit when we take the circle on which the M5-brane is wrapped to infinity  this threefold  for the $\mathbb{Z}_{k}$ case is birationally equivalent to $\widehat{A}_{k-1}\times_{f}\mathbb{C}$ where $\widehat{A}_{k-1}$ is affine $A_{k-1}$ space blown up at $k$ points and it is fibered over $\mathbb{C}$ to obtain the Calabi-Yau threefold.

The affine $A_{k-1}$ space has $k$ $\mathbb{P}^1$'s corresponding to the simple roots. We denote these curve classes by $C_{a}$ with $a=1,2,\cdots, k$. The blow-up introduces $k$ new curve classes which we denote by $M_{a}$ with $a=1,2,\cdots,k$. The class of the elliptic curve $E$ is given by
\bea\label{EC}
E=C_{1}+C_{2}+\cdots+C_{k}\,.
\eea
The complexified K\"ahler parameters associated with these curves are given by the K\"ahler form $\omega$:
\bea\nonumber
t_{a}=\int_{C_{a}}\,\omega\,,\,\,\,\rho=\int_{E}\,\omega\,,\,\,m=\int_{M_{a}}\,\omega\,,\,\,\,\,a=1,2,\cdots,k\,.
\eea
Because of Eq.(\ref{EC}) $\rho$ is given by,
\bea
\rho=t_{1}+t_{2}+\cdots+t_{k}\,.
\eea

The topological string partition function can be calculated using the refined topological vertex \cite{Iqbal:2007ii} and is given by \cite{paper3}:
\bea\label{PF1}
Z_{k}:=\sum_{\vec{\nu}}\prod_{a=0}^{k-1}\Big[(-Q_{a})^{|\nu_{(a)}|}\,G_{\nu_{(a)}\nu_{(a+1)}}\Big]\,,
\eea
where
\bea\nonumber
G_{\nu_{(a)}\nu_{(a+1)}}=\sum_{\mu}(-Q_{m})^{|\mu|}\,C_{\lambda^{t}_{(a)}\,\mu\emptyset}(t, q)\,C_{\lambda_{(a+1)}\mu^{t}\emptyset}(q, t)
\eea
and 
\bea
C_{\lambda\mu\emptyset}(t,q)&=&\Big(\frac{q}{t}\Big)^{\frac{|\lambda|+||\mu||^2-|\mu|}{2}}\,t^{\frac{\kappa(\mu)}{2}}\widetilde{Z}_{\nu}(t,q)\\\nonumber
&&\times\sum_{\eta}\Big(\frac{q}{t}\Big)^{\frac{|\eta|}{2}}\,s_{\lambda^{t}/\eta}(t^{-\rho})\,s_{\mu/\eta}(q^{-\rho})
\eea
is the refined topological vertex (See Appendix A for notation and conventions). The length of the slanted lines in \figref{web2} are all equal to $m$ and we have defined $Q_{m}=e^{2\pi i m}$, similarly the length of the horizontal lines is $T_{a}$ and we have defined $Q_{a}=e^{2\pi i T_{a}}$ such that $t_{a}=log(Q_{a}Q_{m})/2\pi i$ is the distance between the two vertical lines. Using the identity
\bea\nonumber
\sum_{\lambda\,\eta}Q_{\rho}^{|\lambda|}s_{\lambda^{t}/\eta}(x)s_{\lambda/\eta}(y)=\prod_{k=1}^{\infty}
(1-Q_{\rho}^k)^{-1}\prod_{i,j}(1-Q_{\rho}^{k}x_{i}y_{j})\,
\eea
we can calculate the sum in Eq.(\ref{PF1}) to obtain:
\begin{widetext}
\bea\label{PF2}
Z_{k}(\rho,m,{\bf t},\epsilon_{1,2})&=&Z_{1}^{k}\Big(\prod_{n=1}^{\infty}(1-Q_{\rho}^n)\Big)^{k-1}\,\prod_{1\leq a<b\leq k} \,\frac{G_{2}(Q_{ab};\rho,\epsilon_1,\epsilon_2)G_{2}(Q_{\rho}Q_{ab}^{-1};\tau,\epsilon_1,\epsilon_2)}
{G_{2}(Q_{ab}Q_{m}\sqrt{t\,q};\rho,\epsilon_1,\epsilon_2)G_{2}(Q_{ab}\,Q_{m}^{-1}\sqrt{t\,q};\rho,\epsilon_1,\epsilon_2)}\,,\\\nonumber
Z_{1}(\rho,m,\epsilon_{1,2})&=&\prod_{i,j,k=1}^{\infty}\frac{\Big(1-Q_{\rho}^{k-1}\,Q_{m}q^{i-\frac{1}{2}}t^{j-\frac{1}{2}}\Big)
\Big(1-Q_{\rho}^{k}\,Q_{m}^{-1}q^{i-\frac{1}{2}}t^{j-\frac{1}{2}}\Big)}
{\Big(1-Q_{\rho}^{k}\,q^{i}t^{j}\Big)\Big(1-Q_{\rho}^{k}\,q^{i-1}t^{j-1}\Big)}\,,\\\nonumber
G_{2}(x;\rho,\epsilon_{1},\epsilon_{2})&=&\prod_{k,i,j=1}^{\infty}(1-Q_{\rho}^{k-1}q^{i-1}t^{-j+1}x)
(1-Q_{\rho}^{k}q^{i}t^{-j}x^{-1})\,,
\eea
\end{widetext}
where $Q_{ab}=\text{exp}\Big(2\pi i(t_{b}-t_{a})\Big)$.

\subsection{Partition function and Hilbert Scheme of points}

The above partition function that we obtained in Eq.(\ref{PF2}) related to the index of Dirac operator couple to bifundamental matter and can be obtained using a slight generalization of the $\chi_y$ genus. Recall that the $\chi_y$ genus of a manifold $M$ is given by,
\bea
\chi_{y}(M)=\int_{M}\prod_{i}\frac{x_{i}(1-ye^{-x_i})}{1-e^{-x_{i}}}\,,
\eea
where $x_i$ are roots of the Chern polynomial of the tangent bundle of $M$. It can be generalized using a vector bundle $V$ of rank equal to the rank of the tangent bundle to get,
\bea
\chi_{y}(M,V)=\int_{M}\prod_{i}\frac{x_{i}(1-ye^{-w_i})}{1-e^{-x_{i}}}\,,
\eea
where $w_i$ are the roots of the Chern polynomial of $V$. Using $\chi_{y}(M,V)$ we can write the partition function in Eq.(\ref{PF2}) as,
\bea\label{opp}
Z_{k}=\sum_{m_1,m_2,..,m_k}Q_{1}^{m_1}\cdots Q_{k}^{m_{k}}\,\chi_{Q_m}(H_{m_1,..,m_k},V)\,,
\eea
where $H_{m_1,..,m_k}$ is the product of Hilbert scheme of points on $\mathbb{C}^2$,
\bea\nonumber
H_{m_1,..,m_k}=\mbox{Hilb}^{m_1}[\mathbb{C}^2]\times \mbox{Hilb}^{m_2}[\mathbb{C}^2]\times \cdots\times \mbox{Hilb}^{m_k}[\mathbb{C}^2]\,,
\eea
and $V$ is a vector bundle on $H_{m_1,..,m_k}$ whose fiber at a point can be described as follows. Recall that the $\mbox{Hilb}^{m}[\mathbb{C}^2]$ is the resolution of $m$-th symmetric product of $\mathbb{C}^2$ and points in it can be identified with codimension $m$ ideals in the ring $\mathbb{C}[x,y]$ \cite{nakajimabook}. Thus a point in $H_{m_1,..,m_k}$ is a collection of ideals $(I_1,I_2,\cdots,I_k)$ such that $I_{a}\subset \mathbb{C}[x,y]$ and $\mbox{dim}(\mathbb{C}[x,y]/I_{a})=m_a$. The fiber of the vector bundle $V$ over the point $(I_1,I_2,\cdots, I_k)$ is given by,
\bea
\mbox{Ext}^{1}(I_1,I_2)\oplus \mbox{Ext}^{1}(I_2,I_3)\oplus \cdots \oplus \mbox{Ext}^{1}(I_{k},I_{1})\,.
\eea

For the ADE case it is natural to expect that the partition function is still given by an expression of the kind Eq.(\ref{opp}) with manifold being,
\bea
{\cal M}(r_1,m_1)\times {\cal M}(r_2,m_2)\times  \cdots\times {\cal M}(r_k,m_k)\,,
\eea
where ${\cal M}(r,m)$ is the moduli space of $U(r)$ instantons on $\mathbb{C}^2$ with charge $m$ and $r_a$ are the Kac labels of the ADE Dynkin diagram. The description of the bundle in this case is slightly more involved and will be discussed in \cite{KS2}.



\subsection{Modes and Supermultiplets}
From the partition function Eq.(\ref{PF2}) we can determine the contribution of various supermultiplets in five dimensions. Massless particles in five dimensions have little group $Spin(3)=SU(2)$ and the massive particles have little group $Spin(4)=SU(2)_{L}\times SU(2)_{R}$. Fields in the massive multiplet (which is actually $(2,0)$ multiplet of the ${\cal N}=2$ supersymmetry in five dimensions) are in the following representation \cite{susy},
\bea
\left(1,0\right)\oplus 4\left(\frac{1}{2},0\right)\oplus 5\left(0,0\right).
\eea
The five dimensional vector multiplet containing the massless vector field is in the following representation of $Spin(3)$ \cite{susy},
\bea
\left(1\right)\oplus 4\left(\frac{1}{2}\right)\oplus 5\left(0\right).
\eea

\subsubsection{Single M5-brane }
The case of single M5-brane (i.e., $k=1$) was discussed in \cite{Lockhart:2012vp, mstrings}. We discuss it briefly to make comparison later to $k>1$ case. From the plethystic logarithm of the $Z_{1}$ in Eq.(\ref{PF2}) we get,
\bea\nn
F_1&=&Plog(Z_{1})\,,\\\label{kl}
F_{1}+\overline{F_{1}}&=&\underbrace{Q_{m}+Q_{m}^{-1}-\sqrt{\frac{q}{t}}-\sqrt{\frac{t}{q}}}_{massless}+\\\nonumber
&&\sum_{k\in \mathbb{Z},k\neq 0}^{\infty}\underbrace{Q_{\tau}^{k}\Big[(Q_{m}+Q_{m}^{-1})-(\sqrt{q\,t}+\frac{1}{\sqrt{q\,t}})\Big]}_{massive}\,.
\eea
Where in the above equation we have indicated the massless and the KK massive modes.  From Eq.(\ref{kl}) it follows that massless and the massive multiplets are in the following representation of the $SU(2)_{L}\times SU(2)_{R}$,
\bea
\mbox{\it massless}:&&\, \left(0,\frac{1}{2}\right)\oplus 2\left(0,0\right)\,,\\\nn
\mbox{\it massive}:&&\,\,\left(\frac{1}{2},0\right)\oplus 2\left(0,0\right)\,.
\eea
As discussed in \cite{mstrings}  the  contribution of the universal half-hypermultiplet is needed to get the full $Spin(4)$ content \cite{Gopakumar:1998jq}. The half hypermultiplet is $(\frac{1}{2},0)\oplus 2(0,0)$ and tensoring with it and reducing to the diagonal for the massless gives \cite{Lockhart:2012vp,mstrings}:
\bea\nn
\mbox{\it massless}:
&&\left(1\right)\oplus 4\left(\frac{1}{2}\right)\oplus 5\left(0\right),\\\nn
\mbox{\it massive}:&&\left(1,0\right)\oplus 4\left(\frac{1}{2},0\right)\oplus 5\left(0,0\right).
\eea
The above is the six dimensional tensor multiplet reduced on the circle \cite{susy}.

\subsubsection{M5-brane with Transverse Orbifold and Modes}
For the case of M5-brane with transverse orbifold we use $Z_k$ given in Eq.(\ref{PF2}). To determine the spin content of the various modes we consider $F_k+\overline{F_k}$ where $F_k$ is defined by,
\bea\nn
Z_{k}=\mbox{exp}\Big(-\sum_{n=1}^{\infty}\frac{F_{k}(n\rho,nm,n\epsilon_1,n\epsilon_2)}{n(q^{\frac{n}{2}}-q^{-\frac{n}{2}})(t^{\frac{n}{2}}-t^{-\frac{n}{2}})}\Big)
\eea
In \cite{Benvenuti:2016dcs} the form of $F_{k}$ was conjectured by taking into account the Kaluza-Klein charge of the wrapping instanton. Using the identity,
\bea\nonumber
G_{2}(x;\rho,\epsilon_1,-\epsilon_2)=
PE\Big(-\frac{x+x^{-1}Q_{\rho}\,q\,t}{(1-Q_{\rho})(1-q)(1-t)}\Big)\,,
\eea
where $PE$ denotes the plethystic exponential we get ,
\begin{widetext}

\bea\nn
F_{k}+\overline{F_{k}}
&=&k\Big(Q_{m}+Q_{m}^{-1}-\sqrt{\frac{t}{q}}-\sqrt{\frac{q}{t}}\Big)+\sum_{n\in \mathbb{Z},n\neq 0}Q_{\rho}^{n}\Big[k\Big(Q_{m}+Q_{m}^{-1}\Big)-\sqrt{t\,q}-\frac{1}{\sqrt{t\,q}}-(k-1)\Big(\sqrt{\frac{t}{q}}+\sqrt{\frac{q}{t}}\Big)\Big]\\&&+\sum_{1\leq a<b\leq k}\sum_{n\in \mathbb{Z}}Q_{\rho}^{n}(Q_{ab}+Q_{ab}^{-1})\Big(Q_{m}+Q_{m}^{-1}-\sqrt{t\,q}-\frac{1}{\sqrt{t\,q}}\Big)
\eea

\end{widetext}
From the above we see that the massless and the massive multiplets are in the following representations of $SU(2)_{L}\times SU(2)_{R}$,
\bea\nonumber
\text{Massless:}&&\Big[2(0,0)\oplus \,(0,\frac{1}{2})\Big] \oplus (k-1)\Big[2(0,0)\oplus \,(0,\frac{1}{2})\Big]\\\nn
\text{Massive:}&& \Big[2(0,0)\oplus (\frac{1}{2},0)\Big]\,\,\oplus\,\,(k-1)\Big[2(0,0)\oplus \,(0,\frac{1}{2})\Big]\\\nonumber&&\oplus \frac{k(k-1)}{2} \Big[2(0,0)\oplus (\frac{1}{2},0)\Big]
\eea
As discussed for the $k=1$ case we are still missing the contribution of the universal half hypermultiplet associated with the position of the particle in $\mathbb{R}^4$.  Tensoring with the half hypermultiplet $2(0,0)\oplus (\frac{1}{2},0)$ gives,
\begin{widetext}

\bea
\text{Massless:} && \Big[\underbrace{(1)\oplus 4(\frac{1}{2})+5(0)}_{\text{Tensor Multiplet}}\Big]\oplus (k-1)\Big[\underbrace{(1)\oplus 4(\frac{1}{2})+5(0)}_{\text{Massless Vector Multiplet}}\Big]\\\nn
\text{Massive:} && \Big[\underbrace{(1,0)\oplus 4(\frac{1}{2},0)\oplus 5(0,0)}_{\text{Tensor Multiplet}}\Big]\oplus (k-1)\Big[\underbrace{(\frac{1}{2},\frac{1}{2})\oplus 2(0,\frac{1}{2})\oplus 2(\frac{1}{2},0)\oplus 4(0,0)}_{\text{Massive Vector Multiplet}}\Big]\\\nn
&&\oplus \underbrace{\frac{k(k-1)}{2}\Big[(1,0)\oplus 4(\frac{1}{2},0)\oplus 5(0,0)\Big]}_{\text{BPS State for each positive root}}
\eea
\end{widetext}

Thus we see that the supermultiplets correspond to reduction of one 6D tensor multiplet on a circle together with $(k-1)$ vector multiplets coming from the three form field which we can writ as:
\bea
C^{(3)}=\sum_{a=1}^{k-1}A^{a}\wedge \omega^{a}\,,
\eea
where $A^{a}$ is the gauge field of the vector multiplet and $\omega^a$ are the K\"ahler form for the curves corresponding to simple positive roots in the resolved orbifold. 



\section{ADE Orbifolds and Bound States of M-strings}

In the previous sections we discussed the partition function of an M5-brane with transverse $A_N$ geometry. This partition function can be obtained in many different ways using the duality of this brane configuration with Calabi-Yau geometry, 2D gauge theories or 2D sigma models. For the case when the transverse orbifold is of $D$ or $E$ type the topological vertex can not be used since the corresponding Calabi-Yau threefold is not toric. However, by using the fact that holomorphic curves in the geometry are in one to one correspondence with positive roots of the ADE groups we can generalize the partition function to all ADE cases to obtain:
\begin{widetext}
\bea\nn
Z_{G}=Z_{1}^{r+1}\Big(\prod_{n=1}^{\infty}(1-Q_{\rho}^n)\Big)^{r}\,\prod_{\alpha\in \Delta_{+}}\frac{G_{2}(Q_{\alpha}Q_{m}\sqrt{t\,q};\rho,\epsilon_1,-\epsilon_2)G_{2}(Q_{\alpha}\,Q_{m}^{-1}\sqrt{t\,q};\rho,\epsilon_1,-\epsilon_2)}
{G_{2}(Q_{\alpha}\,t;\rho,\epsilon_1,-\epsilon_2)G_{2}(Q_{\alpha}\,q;\rho,\epsilon_1,-\epsilon_2)}\,.
\label{ADEPF}
\eea
\end{widetext}
where $r$ is the rank of the group $G$ and $\Delta_{+}$ is the set of positive real roots corresponding affine ADE group. The parameters $Q_{\alpha}$ and $Q_{\rho}$ are related to the K\"ahler parameters of the resolved orbifold,
\bea\nonumber
Q_{\alpha}=e^{-\int_{C_{\alpha}}\omega}\,,~~~~~~Q_{\rho}=e^{-\int_{C_{\delta}}\omega}\,,
\eea
where $C_{\alpha}$ is the holomorphic curve corresponding to the positive real root $\alpha$ and $C_{\delta}$ is the holomorphic curve corresponding to the imaginary root $\delta$ of the affine ADE group. In the above equation $\omega$ is the complexified K\"ahler form of the resolved orbifold. Since the function $G_2$ itself is a product we see that $Z_G$ is an infinite product over positive roots of the affine $G$. Due to modular transformation properties of the function $G_{2}$ the partition function $Z_{G}$ satisfies a non-perturbative modular transformation \cite{Lockhart:2012vp,KIP},
\bea\nonumber
Z_{G}\Big(-\frac{1}{\rho},\frac{\epsilon_1}{\rho},\frac{\epsilon_2}{\rho}\Big)=\frac{Z_{G}(\rho,\epsilon_1,\epsilon_2)}{Z_{G}(\frac{\rho}{\epsilon_1},-\frac{1}{\epsilon_1},\frac{\epsilon_2}{\epsilon_1})\,Z_{G}(\frac{\rho}{\epsilon_2},\frac{\epsilon_1}{\epsilon_2},-\frac{1}{\epsilon_2})}\,.
\eea
From the partition function in Eq(\ref{ADEPF}) we can calculate the bound states by taking the plethystic logarithm,
\begin{widetext}
\bea
F_G&=&\mbox{Plog}\, Z_{G}\\\nonumber
&=& (r+1)\,F_{1}+r\frac{Q_{\rho}(\sqrt{t\,q}+\frac{1}{\sqrt{t\,q}}-\sqrt{\frac{t}{q}}-\sqrt{\frac{q}{t}})}{1-Q_{\rho}}+\sum_{\alpha\in \Delta_{+}}\frac{(Q_{\alpha}+Q_{\alpha}^{-1}Q_{\rho})(Q_{m}+Q_{m}^{-1}-\sqrt{t\,q}-\frac{1}{\sqrt{q\,t}})}{(1-Q_{\rho})}\,.
\eea
\end{widetext}
Thus we see that there is a single state corresponding to each positive root and there are no bound states corresponding to sum of positive roots. Actually bound states appear when we turn on the momentum along the compact direction. To see this recall that a dual description of this brane system is given by $N=1^{*}$ 5D affine ADE gauge theory. The partition function in Eq.(\ref{ADEPF}) is the perturbative part of partition function of this gauge theory. Including the instanton part requires turning on momentum along the compact direction i.e.., going away from the $\tau\mapsto i\infty$ limit since $\tau$ is the gauge coupling on the gauge theory side and equal to $1/R$ where $R$ is the radius of the compact direction on which the M5-brane is wrapped. Thus if we consider instanton contribution the partition function of the gauge theory becomes
\bea
\widehat{Z}_{G}=Z_{G}\,\Big(1+Q_{\tau}\,I_{1}+\cdots\Big)\,,
\eea
where $I_1$ is the one instanton contribution. Since the one instanton moduli space can be factored 
\bea
\mbox{One instanton moduli space}=\mathbb{C}^2\times {\cal M}_{1}\,,
\eea 
where the factor $\mathbb{C}^2$ gives the position of the instanton in $\mathbb{C}^2$ and ${\cal M}_{1}$ includes the orientation of the instanton within the gauge group. The Hilbert series of ${\cal M}_{1}$ is well known and given by the character of the adjoint representation of the ADE group $\chi_{G,adj}(Q_{i})$ \cite{Benvenuti:2010pq} and the contribution from the $\mathbb{C}^2$ factor is given by $(q^{1/2}-q^{-1/2})^{-1}(t^{1/2}-t^{-1/2})^{-1}$. Thus the partition function becomes:
\bea\nonumber
\widehat{Z}_{G}=Z_{G}\,\Big(1+Q_{\tau}\,\frac{\chi_{G,adj}(Q_{i})}{(q^{\frac{1}{2}}-q^{-\frac{1}{2}})(t^{\frac{1}{2}}-t^{-\frac{1}{2}})}+\cdots\Big)\,.
\eea
Taking the plethystic logarithm to get the contribution of bound states we get:
\bea
\widehat{F}_{G}&=&\mbox{Plog}\widehat{Z}_{G}\\\nonumber
&=&F_{G}+Q_{\tau}\,\frac{\chi_{G,adj}(Q_{i})}{(q^{\frac{1}{2}}-q^{-\frac{1}{2}})(t^{\frac{1}{2}}-t^{-\frac{1}{2}})}+\cdots\,.
\eea
Since the factor $\chi_{G,adj}(Q_{i})$ contains products of factors $Q_{i}$ corresponding to different simple roots therefore we see that we now have bound states corresponding to sum of positive roots. These bound states disappear in the limit $Q_{\tau}\mapsto 0$ and hence these states corresponding to simple roots form bound states only with the help of momentum on the compact circle.

\section{Conclusions}
In this short paper we discussed the partition function of an M5-brane with transverse orbifold. We saw that the partition function calculated using the topological vertex captures the modes of the tensor multiplet corresponding to the M5-brane as well as the modes of the 3-form field on the transverse orbifold. We generalized the partition function to the case of ADE orbifold and argued that bound states of M-strings corresponding to pair of simple roots appear when the M-5brane is wrapped on a circle and momentum is turned on the circle.\\ 

It will be useful to study the dual partition function coming from the worldvolume of the M2-branes for arbitrary ADE orbifold. In the case of A type orbifold the M2-brane theory in the infrared gives a $(0,2)$ sigma model whose target space is the  product of Hilbert scheme of points on $\mathbb{C}^2$. For arbitrary ADE it is expected that the target space space is product of instanton moduli space with ranks equal to the dynkin label of the corresponding ADE diagram \cite{KS2}. 

\section*{Appendix A}
We give some useful formulas in this appendix which have been used in paper. In the text Greek letters $\lambda,\mu,\nu$ have been used to denote partitions of natural numbers and the notation $(i,j)\in \lambda$ represents coordinates of a box in the Young diagram of the partition $\lambda$. The index $i$ labels the parts of the partition and goes from $1$ to $\ell(\lambda)$, where $\ell(\lambda)$ is the number of parts of the partition. If we fix $i$ the index $j$ takes values in the set $\{1,2,\cdots,\lambda_{i}\}$ where $\lambda_i$ is the $i$-th part of the partition. $\lambda^t$ denotes the transpose of the partition which is obtained by reflection of the Young diagram of $\lambda$ in the diagonal.

The refined topological vertex is defined as \cite{HIV, Iqbal:2007ii}:
\bea\label{TVdefinition}
&&C_{\lambda\mu\nu}(t,q)=\Big(\frac{q}{t}\Big)^{\frac{||\mu||^2}{2}}\,t^{\frac{\kappa(\mu)}{2}}\,q^{\frac{||\nu||^2}{2}}\,\widetilde{Z}_{\nu}(t,q)\,\\\nn
&&\times\sum_{\eta}\Big(\frac{q}{t}\Big)^{\frac{|\eta|+|\lambda|-|\mu|}{2}}\,s_{\lambda^t/\eta}(t^{-\rho}q^{-\nu})\,s_{\mu/\eta}(t^{-\nu^t}q^{-\rho})\,,
\eea
where 
\bea
|\mu|&=&\sum_{i=1}^{\ell(\mu)}\mu_i\,,\,\,\,\,||\mu||^2=\sum_{i=1}^{\ell(\mu)}\mu^{2}_i\,\\\nn
\kappa(\mu)&=&||\mu||^2-||\mu^t||^2\,,
\eea
and $\widetilde{Z}_{\nu}(t,q)$ is defined as,
\bea
\widetilde{Z}_{\nu}(t,q)=\prod_{(i,j)\in \nu}\Big(1-q^{\nu_{i}-j}\,t^{\nu^{t}_{j}-i}\Big)^{-1}\,.
\eea
The functions $s_{\lambda/\eta}({\bf x})$ are the skew-Schur functions defined as,
\bea\nonumber
s_{\lambda/\eta}({\bf x})=\sum_{\mu}N^{\lambda}_{\eta\mu}s_{\mu}({\bf x})\,,
\eea
where $s_{\lambda}({\bf x})$ is the Schur function in variables ${\bf x}=\{x_{1},x_{2},\cdots\}$ and $N^{\lambda}_{\eta\mu}$ are the Littlewood-Richardson coefficients.

The refined vertex is a function of two parameters $q$ and $t$. For $q=t$ it reduces to the usual topological vertex \cite{Aganagic:2003db} with $g_{s}=\mbox{ln}q$ being the topological string coupling constant. The parameters $q$ and $t$ are related to the Omega background parameters $(\epsilon_1,\epsilon_2)$ \cite{Nekrasov:2002qd} as follows,
\bea\nn
(q,t)=(e^{i\epsilon_1},e^{-i\epsilon_2})\,,
\eea
so that the unrefined case corresponds to $\epsilon_{1}+\epsilon_{2}=0$.

\end{document}